\documentclass[openacc]{rstransa}

\usepackage{caption}
\usepackage{subcaption}

\titlehead{Research}

\begin{document}

\title{Investigating Ultra-Large Large-Scale Structures: Potential
  Implications for Cosmology}

\author{
  A. M. Lopez$^{1}$, R. G. Clowes$^{1}$ and G. M. Williger$^{2}$}

\address{$^{1}$Jeremiah Horrocks Institute, University of Central Lancashire,
  Preston, PR1 2HE, United Kingdom\\
  $^{2}$Department of Physics and Astronomy, University of Louisville,
  Louisville, KY 40292, USA }

  \subject{cosmology, standard cosmological model, extragalactic astrophysics}


\corres{Alexia M. Lopez\\
  \email{missalexia.lopez@gmail.com}}

  \begin{abstract}

Large-scale structure (LSS) studies in cosmology map and analyse matter in the Universe on the largest
scales. Understanding the LSS can provide observational support for the
Cosmological Principle (CP) and the Standard Cosmological Model
($\Lambda$CDM).

In recent years, many discoveries have been made of LSSs that are so large
that they become difficult to understand within $\Lambda$CDM.  Reasons for
this are: they potentially challenge the CP, (i.e.\ the scale of
homogeneity); and their formation and origin are not fully understood.

In this article we review two recent LSS discoveries:
the Giant Arc (GA, $\sim 1$~Gpc) and the Big Ring (BR, $\sim 400$~Mpc). Both structures are in the same cosmological
neighbourhood --- at the same redshift $z \sim 0.8$ and with a separation on
the sky of only $\sim 12^\circ$. Both structures exceed the often-cited scale
of homogeneity (Yadav+ 2010), so individually and together, these two
intriguing structures raise more questions for the validity of the CP and
potentially hint at new physics beyond the Standard Model.

The GA and BR were discovered using a novel method of mapping faint matter at intermediate
redshifts, interpreted from the Mg~{\sc II} absorption doublets seen in the
spectra of background quasars.


\end{abstract}


\begin{fmtext}

\end{fmtext}


\maketitle

\section{Introduction}

Large-scale structure (LSS) studies are motivated by the need for
observational data to confirm the predictions of the Standard Cosmological
Model ($\Lambda$CDM). In particular, from studying the LSS of matter on the
very largest scales, one can learn about the growth of cosmic structure
\cite{Huterer2023} and about the Universe's dynamical history
\cite{ChowMartinez2014}, thus allowing comparison with
$\Lambda$CDM. Furthermore, and of some current interest, LSS can test the
assumption of large-scale homogeneity, which is a fundamental aspect of the
Cosmological Principle (CP) and hence of the theoretical framework in
cosmology.

Unfortunately, the CP lacks a precise and agreed definition. Different
interpretations can be encountered in the cosmological literature and across
the history of cosmology. The details can be vague. For example, the textbook
version of the CP might say that the Universe on large scales is homogeneous
and isotropic. However, what those large scales might be is often not clearly
specified, and, indeed, the expectation of what is plausible seems to have
increased by at least a factor of ten over the years.

Consider the following three interpretations of the CP, and specifically what
is meant by homogeneity. (i) There exists some large scale, known as the
scale of homogeneity on which the Universe can be smoothed and the
distribution of matter would then be well represented by a stationary random
process, e.g., \cite{Peebles1993, Yadav2010, Aluri2023}.  (ii) The power
spectrum suggests that there can always be some large scale at which
statistically-significant deviations might be found in the matter
distribution, but such deviations on large scales should be rare, e.g.,
\cite{Park2012, ChowMartinez2014}. (iii) There should be similarity
everywhere (maximally symmetric). Any observed large-scale structuring
indicates that the scale of homogeneity, if it exists, must then be larger
than these scales. The occurrence of a particular LSS, even the largest
known, does not imply that the \emph{probability} of finding a comparable LSS
elsewhere is any different. Points (ii) and (iii) could be contradictory,
given that point (ii) suggests that the largest structures should be rare,
and point (iii) suggests that the largest structures need not be a problem
for the CP if their probabilities are homogeneous and isotropic. (How would
we know that?) A useful overview of the various interpretations of the CP can
be found in \cite{Schwarz2010}.

The diversity in the interpretation of homogeneity in the CP has led to
differing conclusions on whether the observed matter supports a homogeneous
Universe. For example, there have been claims that large-survey analysis
supports homogeneity in luminous red galaxies and in quasars
\cite{Andrade2022, Goncalves2021, Goncalves2018}.  However, the accumulating
set of ultra-large LSS (uLSS)
\footnote{We have introduced the new term `ultra-large LSS' (uLSS) to denote
those structures that exceed the Yadav estimated $\sim 370$~Mpc upper limit
to the scale of homogeneity \cite{Yadav2010}.} discoveries might indicate
that homogeneity is not supported. For a recent list of the largest LSSs that
appear to extend beyond Yadav's \cite{Yadav2010} estimated $\sim 370$~Mpc
upper limit to the scale of homogeneity, see Table~$1$ in \cite{Lopez2022}.

Perhaps individual uLSS discoveries can instead be explained by
appealing to extreme-value statistics in some form. For example,
\cite{Park2012} and \cite{Nadathur2013} address the Sloan Great Wall (SGW)
and the Huge Large Quasar Group (Huge-LQG) with mock catalogues and random
catalogues respectively, finding that mock / random structures of comparable
size and overdensity may be readily reproduced. Somewhat differently for the
Huge-LQG, \cite{Marinello2016} used the Horizon Run 2 cosmological simulation
and extreme-value analysis to show that the Huge-LQG is compatible with the
standard $\Lambda$CDM model if it should happen to be the largest such
structure in a volume over five times larger. Similarly, for the SGW,
\cite{Park2012} found that, while structures comparable to the size and
overdensity of the SGW were reproduced in their simulations, they were always
in the top six largest and richest structures detected in the $200$ mock
samples.

These results from \cite{Park2012} and \cite{Marinello2016} suggest that, in
the matter of the compatibility of uLSSs with the $\Lambda$CDM model, the
\emph{accumulated} set of uLSSs might be of more importance than any one
individual structure.
  
\begin{figure}[!h]
\centering\includegraphics[width=2.5in]{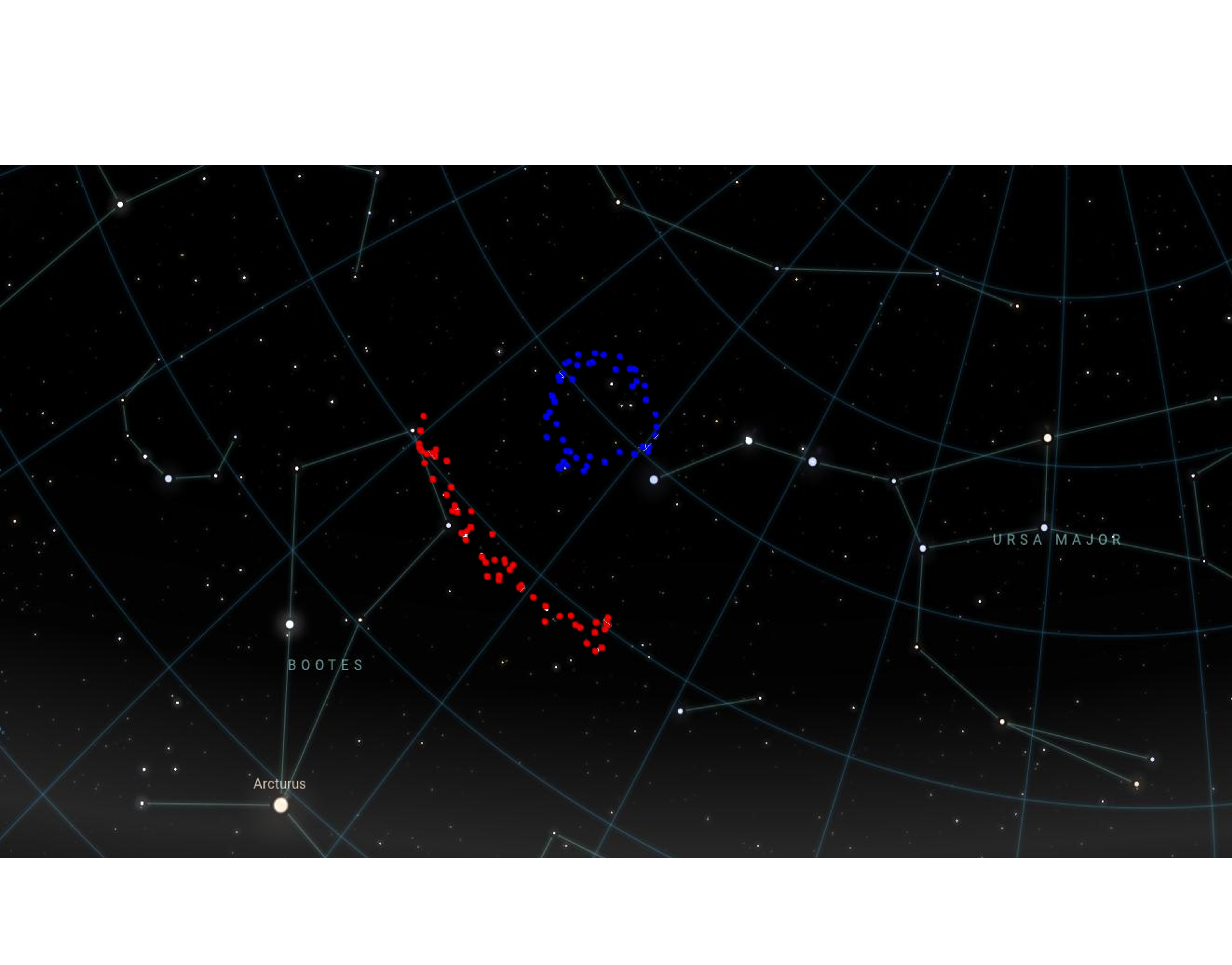}
\caption{An approximate projection of the GA and BR visually-identified
  absorber members superimposed onto an image of the night sky taken from
  Stellarium. This figure gives an impression of the scale of the two uLSSs,
  their position on the sky and their proximity.}
\label{fig:BR_and_GA_night_sky}
\end{figure}

In this review article we discuss two intriguing uLSS discoveries, the Giant
Arc (GA) \cite{Lopez2022} and the Big Ring (BR) \cite{Lopez2024}. See Figure~\ref{fig:BR_and_GA_night_sky} 
for an artistic impression of both of these structures on the sky. We
summarise their method of discovery and statistical analysis, describe their
observed properties, and comment or speculate on their possible origins. The GA
and the BR are in the same cosmological neighbourhood, at $z \sim 0.8$ and
separated by only $\sim 12^\circ$ on the sky. Individually and together, they
exceed the Yadav $370$~Mpc scale, and thus may at least challenge some
interpretations of the CP. 
(Note that the word ‘challenge’ is not synonymous with ‘contradict’, but it  does imply something to be investigated further.) 
Their sizes and morphologies appear to be hard to
explain in the standard $\Lambda$CDM model. Quite possibly, new developments
in cosmology will follow from the continued investigation of intriguing and
unexpected anomalies such as these.

 \section{The Mg~{\sc II} Method}

 Observations for studying LSS in the Universe could be broadly divided into
 two categories: (a) observations of low-luminosity objects at low to
 intermediate redshifts (e.g., \cite{Geller1989, Gott2005, Pomarede2020});
 and (b) observations of high-luminosity objects at intermediate to high
 redshifts (e.g., \cite{Webster1982, Crampton1987, Crampton1989, Clowes2013,
   Horvath2014, Horvath2015, Balazs2015, Balazs2018}). There are benefits and
 challenges to both categories. In category (a), photometric redshifts, for
 very large numbers of objects, are commonplace, but the large redshift
 errors can lead to blurring of structures along the line of
 sight. Spectroscopic redshifts, generally for smaller numbers of objects,
 are more demanding of telescope time, and are more likely to feature in
 category (b). The redshift errors are then smaller but may still be
 associated with some blurring.
 
A novel method for analysing LSS is to infer the low-luminosity matter at
intermediate redshifts from the presence of sharp metal absorption lines in
the spectra of high-redshift quasars.  The Mg~{\sc II} doublet, specifically,
arises from low-ionised metal-enriched gas which is well-known to trace star
formation regions \cite{Bergeron1988, Steidel1992, Churchill2005,
  Kacprzak2008, Barnes2014}.  From many quasar observations covering a large
area of sky, such as the SDSS footprint or the recent DESI survey footprint
(at the time of writing this is not yet publicly available), one can then map
the inferred, low-luminosity, intervening matter at intermediate redshifts
and learn about the LSS.

The sources of data for the Mg~{\sc II} method are these. The first-order
data are the spectroscopic quasar observations from the Sloan Digital Sky
Survey (SDSS) quasar catalogues.  For previous work (including the discovery
and analysis of the GA) we used the `cleaned' quasar catalogues DR7QSO
\cite{Schneider2010} and DR12Q \cite{Paris2017}. For recent work (including
the discovery and analysis of the BR) we used the newer `cleaned' DR16Q
quasar catalogue \cite{Lyke2020}.  The second-order data are the
corresponding Mg~{\sc II} catalogues from independent authors. For the
Mg~{\sc II} absorber catalogues corresponding to DR7QSO and DR12Q we
downloaded the Zhu and M\'enard (Z\&M) \cite{Zhu2013} data, and for the
absorber catalogues corresponding to DR16Q we downloaded the Anand et
al. (Anand21) \cite{Anand2021} data.

\section{Statistical Analysis}

In this section we summarise the statistical analyses performed on the GA and
BR to assess them, sub-divided by the different statistical tests used. Full
details can been found in the respective papers.

\subsection{Power Spectrum Analysis}

\begin{figure}[!h]
\centering\includegraphics[width=3in]{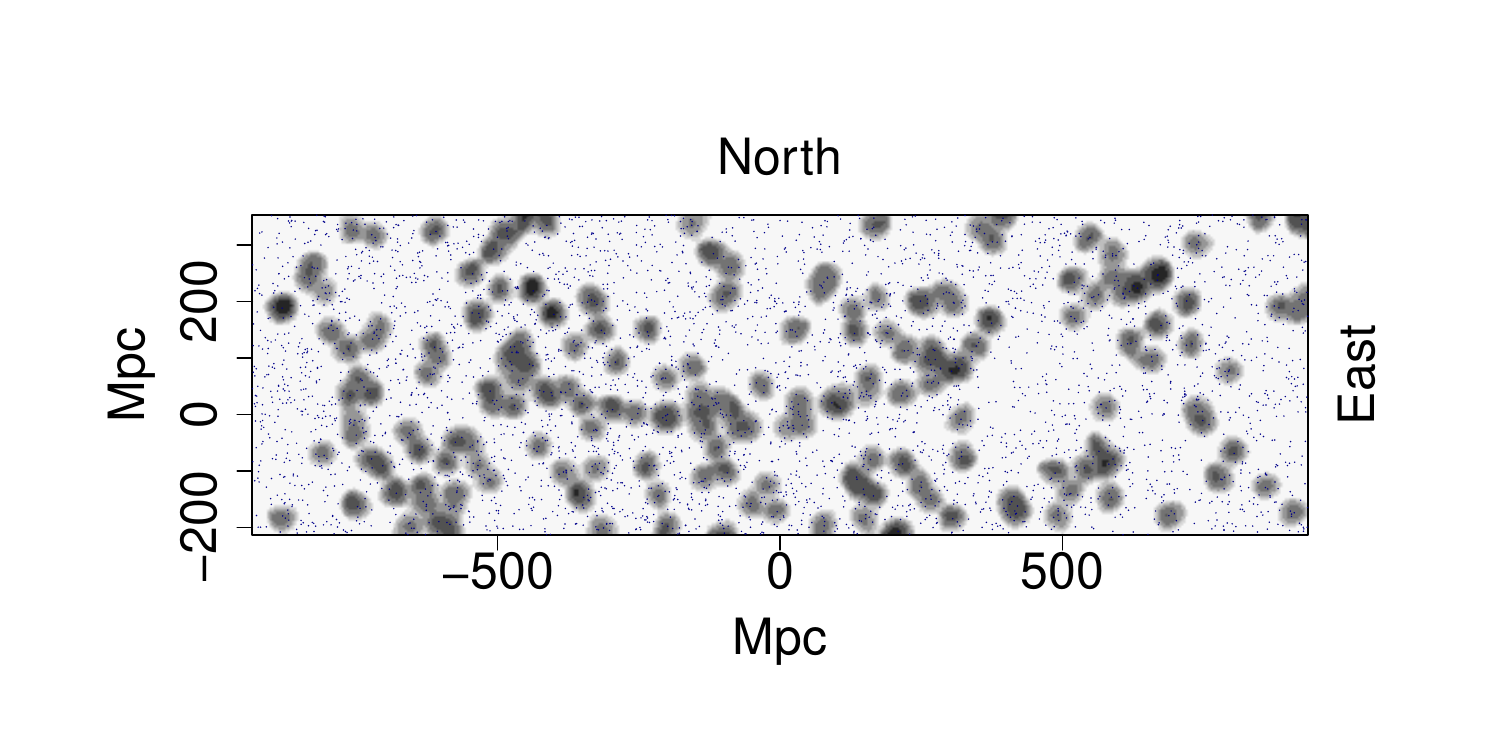}
\caption{The tangent-plane distribution of Mg~{\sc II} absorbers centred on
  the GA in the redshift slice $z = 0.802 \pm 0.060$ using the Z\&M data. The
  grey contours, increasing by a factor of two, represent the density
  distribution of the absorbers which have been smoothed using a Gaussian
  kernel of $\sigma = 11$~Mpc, and flat-fielded with respect to the
  distribution of background probes (quasars). The dark blue dots represent
  the background probes. This figure corresponds to Figure~9a of the GA
  paper.}
\label{fig:GA_field_for_PSA}
\end{figure}

The 2D Power Spectrum Analysis (2D PSA) \cite{Webster1976a, Webster1976b} is a powerful
statistical tool for detecting clustering of sources in a rectangular field
using Fourier methods. The 2D PSA has the power to detect the scale of
clustering as well as its statistical significance on that scale. We applied
this test to the field containing the GA. Given that the (likely) dominant
feature of clustering in the GA field was the GA itself, we reduced the
typical Mg~{\sc II} image size, mostly along the north-south axis (Figure
\ref{fig:GA_field_for_PSA} ), before applying the 2D PSA.

\begin{figure}[!h]
\centering\includegraphics[width=3in]{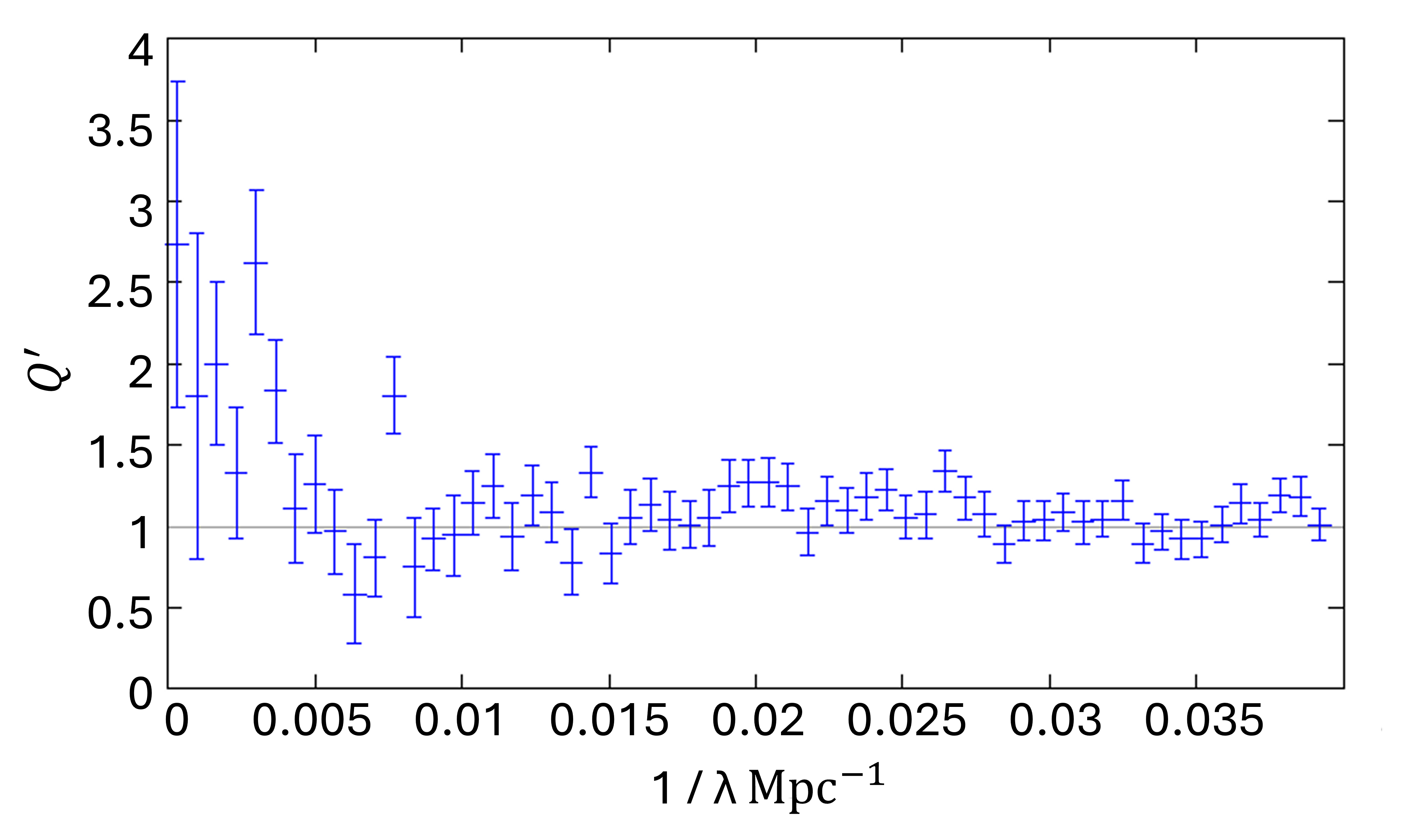}
\caption{The 2D PSA test statistic $Q'$ as a function of clustering scale
  $1/\lambda$ with $\lambda$ in Mpc. The bin size is $6.7 \times
  10^{-4}$~Mpc$^{-1}$ and the error bars are $\pm \sigma$.  The horizontal
  line $Q' =1$ indicates the expectation value in the case of no clustering.
  The (six) high points towards the left of the plot allow a clustering scale
  of $\lambda_c \sim 270$~Mpc to be identified. The final PSA statistic $Q$
  for this scale $\lambda_c$ corresponds to a detection of clustering at a
  significance of $4.8\sigma$. This figure corresponds to Figure~13 in the GA
  paper.}
\label{fig:PSA_on_GA}
\end{figure}


In Figure~\ref{fig:PSA_on_GA} we are showing the 2D PSA statistic $Q'$
corresponding to the GA field seen in Figure~\ref{fig:GA_field_for_PSA}. The
(six) high points towards the left of the figure allow a clustering scale
of $\lambda_c \sim 270$~Mpc to be identified. The final PSA statistic $Q$
for this scale $\lambda_c$ corresponds to a detection of clustering at a
significance of $4.8\sigma$. Given
the clustering scale of $\lambda_c \sim 270$~Mpc, this is likely detecting
the width of the GA as perceived along its length.

\subsection{Cuzick and Edwards Test}

The Cuzick and Edwards (CE) test \cite{Cuzick1990} is a case-control $k$ (or $q$ here)
nearest-neighbours algorithm, originally intended to assess geographical,
spatial clustering of medical illnesses in inhomogeneous populations. Given
the complications of the data we work with --- the possible inhomogeneity of
the quasars (the probes) available for detecting intervening Mg~{\sc II}
absorbers --- we applied this test to both the GA and the BR.
    
Cuzick and Edwards found the test to be most powerful when the ratio of
controls to cases is between $4$ and $6$. However, for the GA field there
were $\sim 20$ times as many probes (controls) as Mg~{\sc II} absorbers
(cases), and for the BR field there were $\sim 50$ times as many probes as
Mg~{\sc II} absorbers. For the CE test calculation, in both cases of the GA and BR, 
a subset of probes was randomly selected to
reduce the ratio of controls to cases to $5:1$, and this procedure was then 
repeated $100$ times. Within each run of randomly selected probes,
the CE test was run with $2000$ simulations. 
    
\begin{figure*}
\centering
\begin{subfigure}[b]{0.475\textwidth}
\centering
\includegraphics[width=\textwidth]{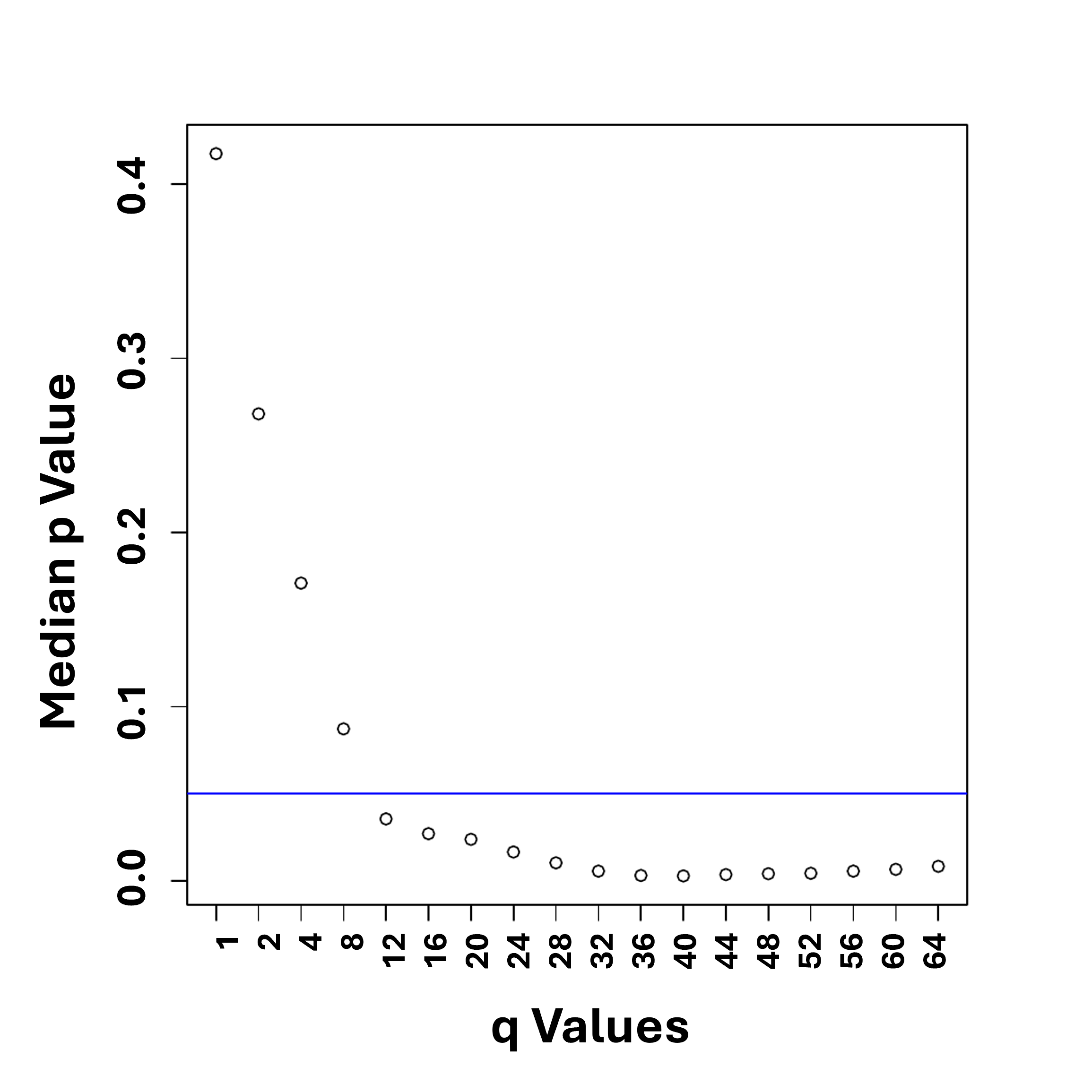}
\caption{}
\label{fig:CE_on_GA}
\end{subfigure}
\hfill
\begin{subfigure}[b]{0.475\textwidth}
\centering
\includegraphics[width=\textwidth]{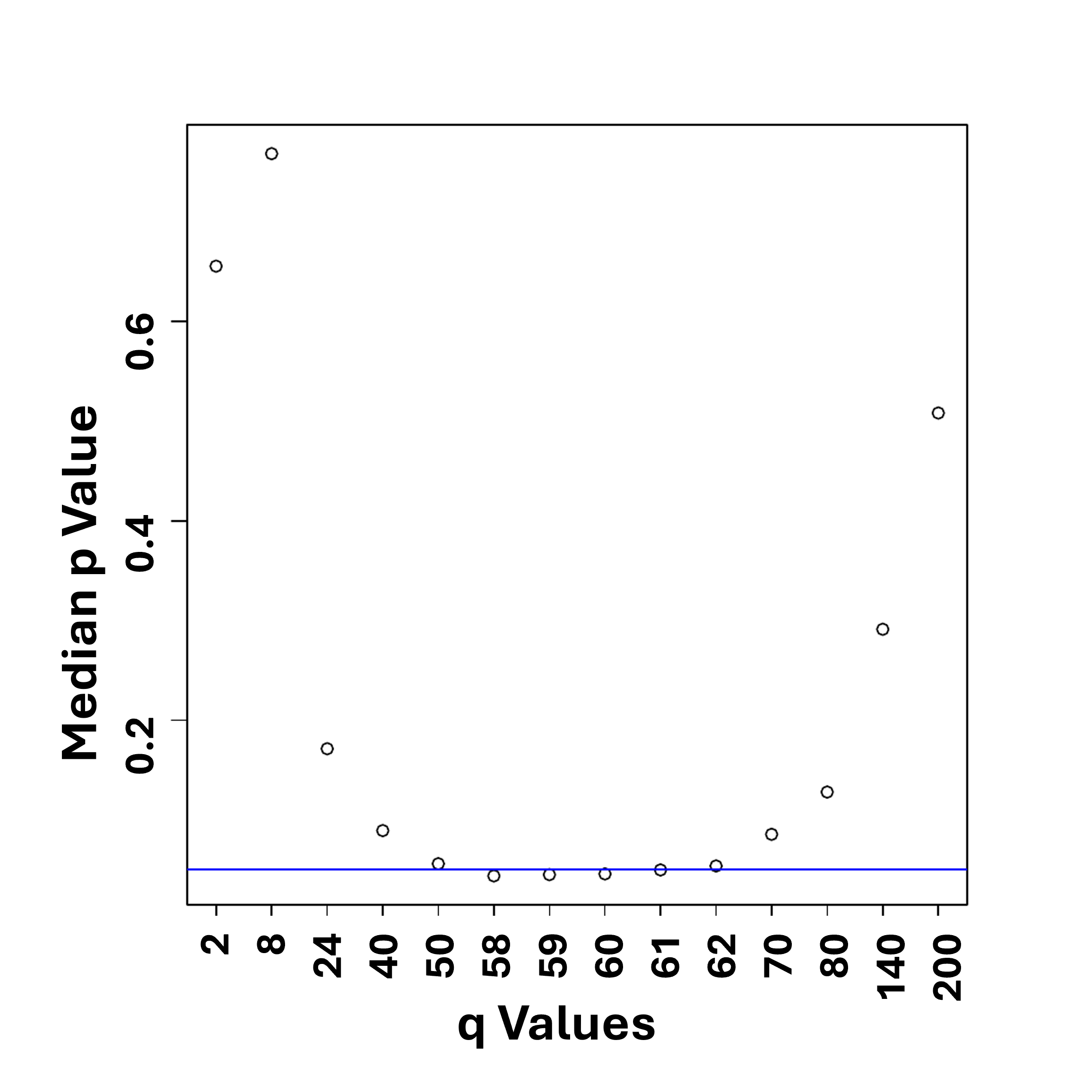}
\caption{}
\label{fig:CE_on_BR}
\end{subfigure}
\caption{\small (a) The median $p$-value over $100$ runs of $2000$
  simulations as a function of chosen $q$ value. These results correspond to
  the GA field seen in Figure~\ref{fig:GA_field_for_PSA}. The $p$-value
  reaches a minimum of $p=0.0027$ (corresponding to a $3 \sigma$ detection)
  at $q=40$. These results are based on data from the older Z\&M
  database. The $x$-axis labels are: $1, 2, 4, 8, 12, 16, 20, 24, 28, 32, 36, 40, 44, 48, 52, 56, 60, 64$. (b) The median $p$-value over $100$ runs of $2000$ simulations as
  a function of chosen $q$ value. These results correspond to the BR field
  seen in Figure~\ref{fig:BR_field_for_CE}. The $p$-value reaches a minimum
  of $p=0.043$ at $q=58$. These results are based on data from the Anand21
  database. The $x$-axis labels are $2, 8, 24, 40, 50, 58, 60, 61, 62, 70, 80, 140, 200$. }
\label{figs:CE}
\end{figure*}

\begin{figure}[!h]
\centering\includegraphics[width=3in]{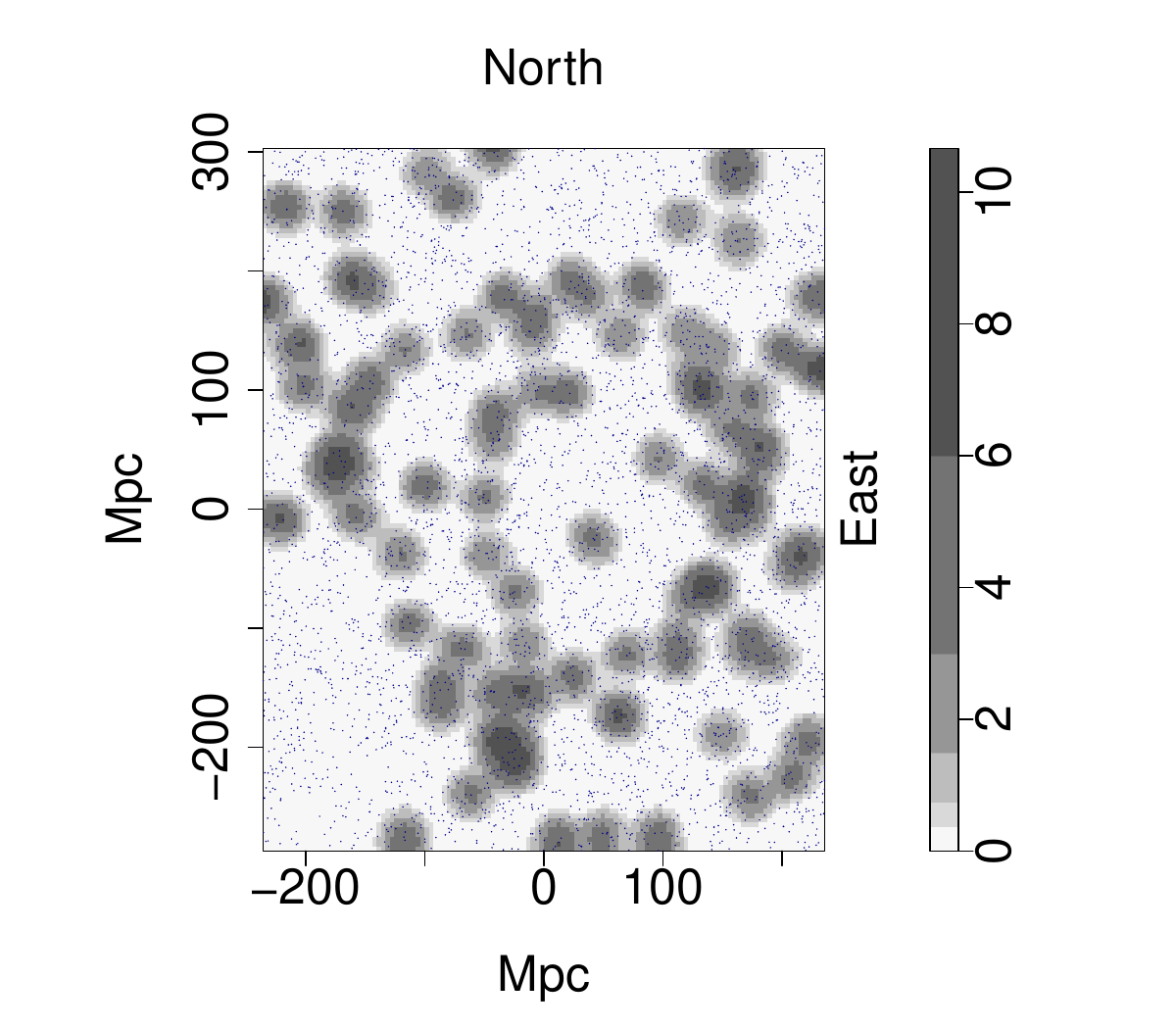}
\caption{The tangent-plane distribution of Mg~{\sc II} absorbers centred on
  the BR in a reduced field size in the redshift slice $z=0.802 \pm 0.060$
  using the Anand21 data. The grey contours
  represent the density distribution of the absorbers which have been
  smoothed using a Gaussian kernel of $\sigma = 11$~Mpc, and flat-fielded
  with respect to the distribution of background probes (quasars). The key
  on the right-hand side shows the relative density of the Mg~{\sc II}
  absorbers, which are increasing by a factor of two.  The dark
  blue dots represent the background probes. }
\label{fig:BR_field_for_CE}
\end{figure}

For the GA field, seen in Figure~\ref{fig:GA_field_for_PSA}, the CE test
found significant $p = 0.0027$ field clustering at $q=40$ (see
Figure~\ref{fig:CE_on_GA}), corresponding to a significance of $3.0 \sigma$.
For the (reduced) BR field seen in Figure~\ref{fig:BR_field_for_CE}, the CE
test did not find conclusive significant clustering. In
Figure~\ref{fig:CE_on_BR} the $p$-value drops below $0.05$ (indicated by the
blue, horizontal line), reaching a minimum of $p=0.043$ ($1.7 \sigma$) at
$q=58$. 

\subsection{FilFinder algorithm}

\begin{figure*}
\centering
\begin{subfigure}[b]{0.475\textwidth}
\centering
\includegraphics[width=\textwidth]{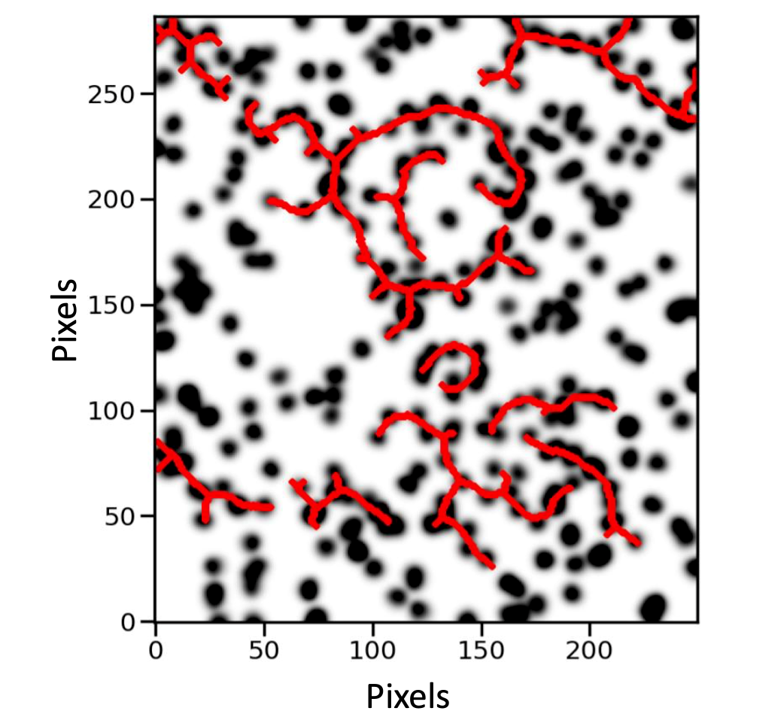}
\caption{}
\label{fig:FilFinder_on_BR}
\end{subfigure}
\hfill
\begin{subfigure}[b]{0.475\textwidth}
\centering
\includegraphics[width=\textwidth]{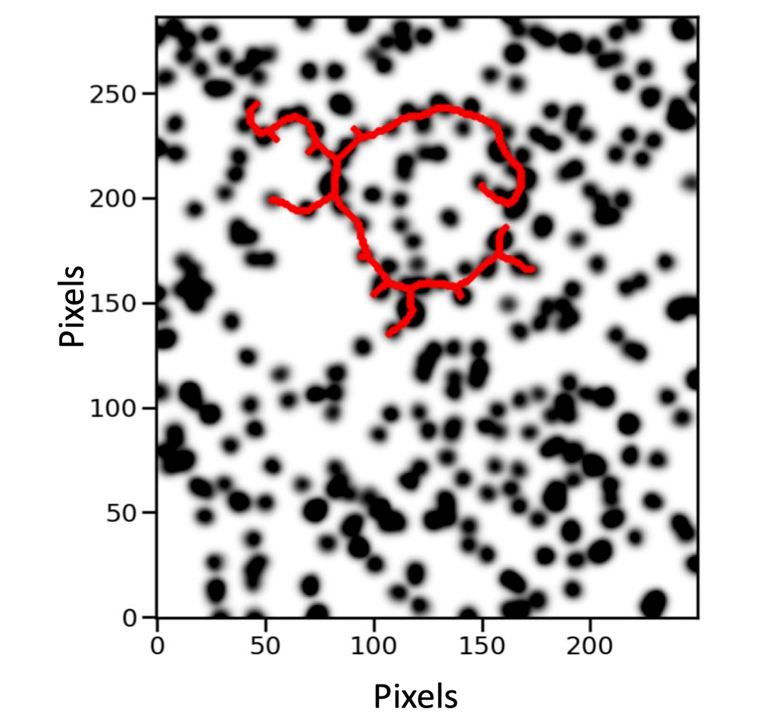}
\caption{}
\label{fig:FilFinder_on_BR_2}
\end{subfigure}
\caption{\small (a) Filaments, shown in red, identified by the FilFinder
  algorithm in the BR field. (b) Starting from (a), the FilFinder size
  threshold parameter was incrementally increased, leaving only the filament
  that traces the absorbers belonging to the BR. The Mg~{\sc II} data are
  from the Anand21 database. }
\label{figs:FilFinder}
\end{figure*}

The FilFinder algorithm \cite{Koch2015} is a 2D filament identification tool,
originally designed to trace filamentary structure in small gaseous regions,
such as the ISM and star-formation regions \cite{Mookerjea2023, Zhang2023,
  Meidt2023}.  We applied it to the field containing the BR to objectively
trace the longest and most-connected filaments in the field.
Figure~\ref{fig:FilFinder_on_BR} shows the FilFinder algorithm applied to the
BR field (for details of the parameter choices, see the original discovery
paper).  By incrementally increasing the size threshold of the algorithm, it
removes the smaller and less connected filaments (see
Figure~\ref{fig:FilFinder_on_BR_2}).  In doing so, only the filament tracing
the BR remained. Applying the FilFinder algorithm to the BR field shows that
the BR is the most connected and largest filament in the field.
    
\subsection{Single Linkage Hierarchical Clustering Algorithm}

The Single-Linkage Hierarchical Clustering (SLHC) algorithm is equivalent to
a Minimal Spanning Tree (MST) when the MST is separated at a specified linkage
scale.  When applied to 3D spatial data the SLHC algorithm determines the
candidate (algorithmic) `structures' within the dataset.  The linkage scale
will determine the set of data points included in a candidate
structure. Naturally, the choice of linkage scale must be related to the
field density: in a high-density region, the chosen linkage scale should be
smaller than in a low-density region.  We adopt the GA field as a standard
field (linkage scale $s_0$, density $\rho_0$) and use the relation
$s = (\rho_0 / \rho)^{1/3}s_0$ to obtain the linkage scale $s$ for any other
field of density $\rho$.

The SLHC algorithm identified the majority of the absorbers belonging to the
visually-identified GA (using the older Z\&M database) in two agglomerations:
a large, statistically-significant portion and a smaller, not
statistically-significant portion, which were named GA-main and GA-sub,
respectively. The two GA agglomerations visually overlapped on the sky,
indicating that they could belong to the same candidate structure if given a
more complete survey coverage. (Recall that the background quasars are
responsible for the detection of intervening Mg~{\sc II} systems.) The
statistical significance of GA-main was then calculated using the Convex Hull
of Member Spheres (see originally \cite{Clowes2012} and the corresponding GA
or BR paper for more details). The CHMS statistical significance of GA-main
was $4.53 \sigma$.  The overdensity of GA-main was calculated both with the
CHMS method and the Pilipenko MST-overdensity method \cite{Pilipenko2007}.
The CHMS-overdensity and MST-overdensity for GA-main were $\delta \rho_{CHMS}
/ \rho = 0.9 \pm 0.6$ and $\delta \rho_{MST} / \rho = 1.3 \pm 0.3$
respectively. Clearly, the larger error estimates from the CHMS-overdensity
indicate that those results should be taken with caution. A mass excess was
estimated from the MST-overdensity by assuming that $\delta_n = \delta_m$,
where $\delta_n$ is the MST-overdensity and $\delta_m$ is the mass
overdensity. We took the critical density of the Universe to be $9.2 \times
10^{-27}$~kg~m$^{-3}$. The estimated mass excess for GA-main was $1.8 \times
10^{18}$M$_{\odot}$.

\begin{figure}[!h]
\centering\includegraphics[width=2.5in]{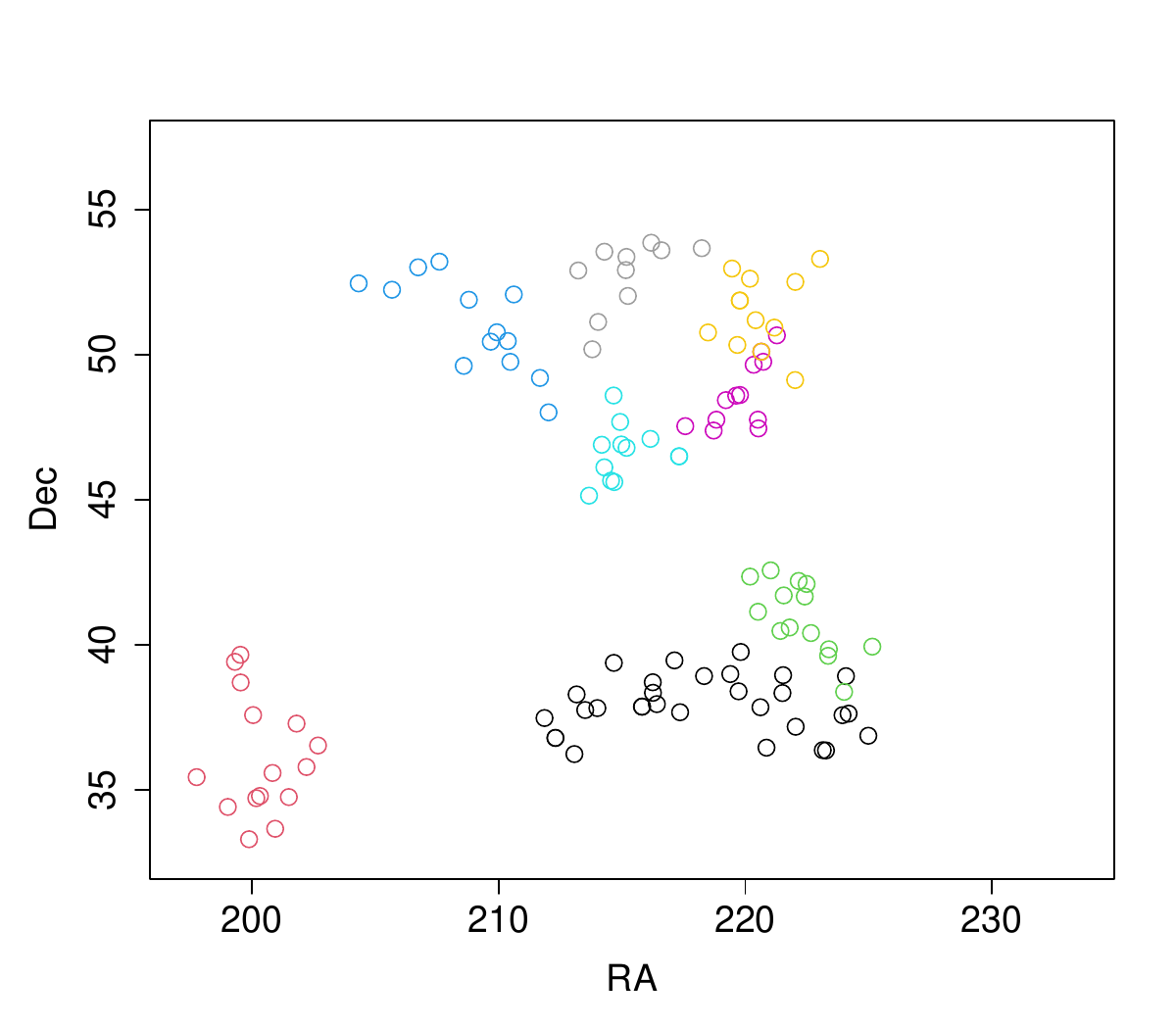}
\caption{Eight of the $10$ highest membership candidate structures identified
  by the SLHC algorithm applied to the BR field in the redshift slice
  $z=0.802 \pm 0.060$. The colours represent the memberships which are
  ordered from high to low in the following way: black, red, green, blue,
  turquoise, pink, yellow, grey. In this figure, only the black points,
  representing absorbers belonging to the GA, are statistically
  significant. Data points are from the Anand21 database.}
\label{fig:BR_SLHC}
\end{figure}

\begin{figure}[!h]
\centering\includegraphics[width=2.5in]{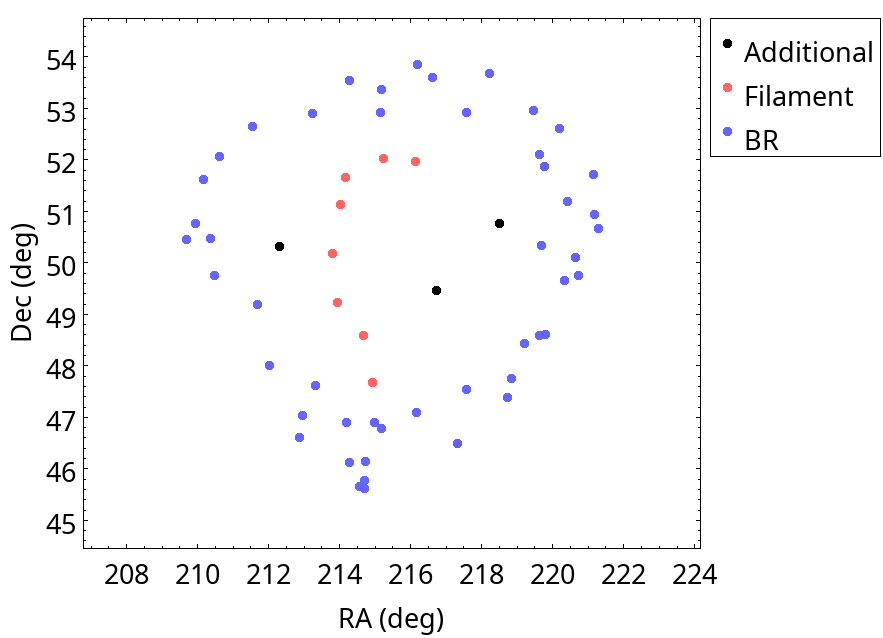}
\caption{The visually-identified BR shown in blue points. The orange points
  mark an interesting filament running through the BR and the black
  points are the additional absorbers encompassed by the BR. Data points are
  from the Anand21 database.}
\label{fig:BR_visual}
\end{figure}

The SLHC algorithm applied to the BR field identified the majority of the
absorbers belonging to the visually-identified BR (using the Anand21
database) in five, adjacent or overlapping, candidate structures,
which, individually, were not statistically significant (see
Figure~\ref{fig:BR_SLHC}, also see Figure~\ref{fig:BR_visual} for our
selection of the visually-identified BR). Similar to what was seen with the
SLHC algorithms on the GA field, the individual candidate structures in the
BR appear connected on the sky, indicating that they could belong to the same
structure. Additionally, the results from the FilFinder add further support
to the connectivity of the BR.  We then apply the CHMS and MST significance
tests to four versions of the BR: those identified by the FilFinder algorithm
(FilFinder-identified); those identified by the SLHC in five candidate
structures (SLHC-identified); the visually-identified absorbers making the
circumference of the BR (BR-only); the visually-identified absorbers of the
BR and everything contained within the BR (BR-all).
Table~\ref{tab:BR_significances} presents the results of the CHMS and MST
significances applied to the four versions of the BR.

\begin{table}[!h]
\caption{The SLHC and MST significance results applied to the four versions
  of the BR from the: SLHC-identified absorbers, visually-identified (BR-all
  and BR-only) absorbers and FilFinder-identified absorbers. These results
  are based on data from the Anand21 database.}
\label{tab:BR_significances}
\begin{tabular}{llllll}
\hline
& SLHC ($\sigma$)& Visual BR-all ($\sigma$)& Visual BR-only ($\sigma$)& FilFinder ($\sigma$)& {\bf Mean} ($\sigma$)\\ 
CHMS & $3.6$ & $5.2$ & $3.3$ & $2.5$ & $3.7 \pm 1.1$\\ 
MST  & $4.7$ & $4.1$ & $4.0$ & $3.6$ & $4.1 \pm 0.5$\\\hline
\end{tabular}
\end{table}

In Table~\ref{tab:BR_significances}, looking across the four versions in the
CHMS significance calculations, the likely upper-limit estimate is from the
visually-identified BR-all absorbers, and the lower-limit estimate is from
the FilFinder-identified absorbers. This variation is expected, as the CHMS
draws a unique volume around the absorber members, and empty volumes existing
between absorbers will also be included. So, with large, empty volumes
incorporated by the CHMS algorithm, the volume of a filamentary-type
structure will be overestimated, and the significance of the unique CHMS
volume will be correspondingly underestimated. For example, compare the
filament-identified absorbers in Figure~\ref{fig:FilFinder_on_BR_2} with the
visually-identified absorbers in Figure~\ref{fig:BR_visual}. The
filament-identified absorbers envelop a large, empty region in the middle, as
well as creating additional empty regions from the additional spurs to the
north-west, whereas the visually-identified absorbers, including all of those
absorbers contained within the visually-identified BR, have fewer empty
regions. The differences are then reflected in the CHMS significance
estimates. Ideally, the CHMS algorithm is more appropriately used for clumpy,
globular-type structures, and is less applicable to filamentary-type
structures.

In contrast, in Table~\ref{tab:BR_significances} the MST-significance values
are less varying than the CHMS-significance values, as was also the case for
the GA (see above). The CHMS algorithm relies on the unique volume
encompassing the absorber members, whereas the MST calculation relies on the
mean MST edge length between absorber members, which might be better suited
for filamentary structures than the CHMS.

\section{Observational Properties}

In this section we give examples of and review some of the observational
properties of the GA and the BR; further details can be found in the
respective papers.

\subsection{Independent Corroboration}

We used the SDSS DR16Q quasars and the DESI cluster catalogue from
\cite{Zou2021} to provide independent corroboration of the Mg~{\sc II}
structures.

\begin{figure*}
\centering
\begin{subfigure}[b]{0.475\textwidth}
\centering
\includegraphics[width=\textwidth]{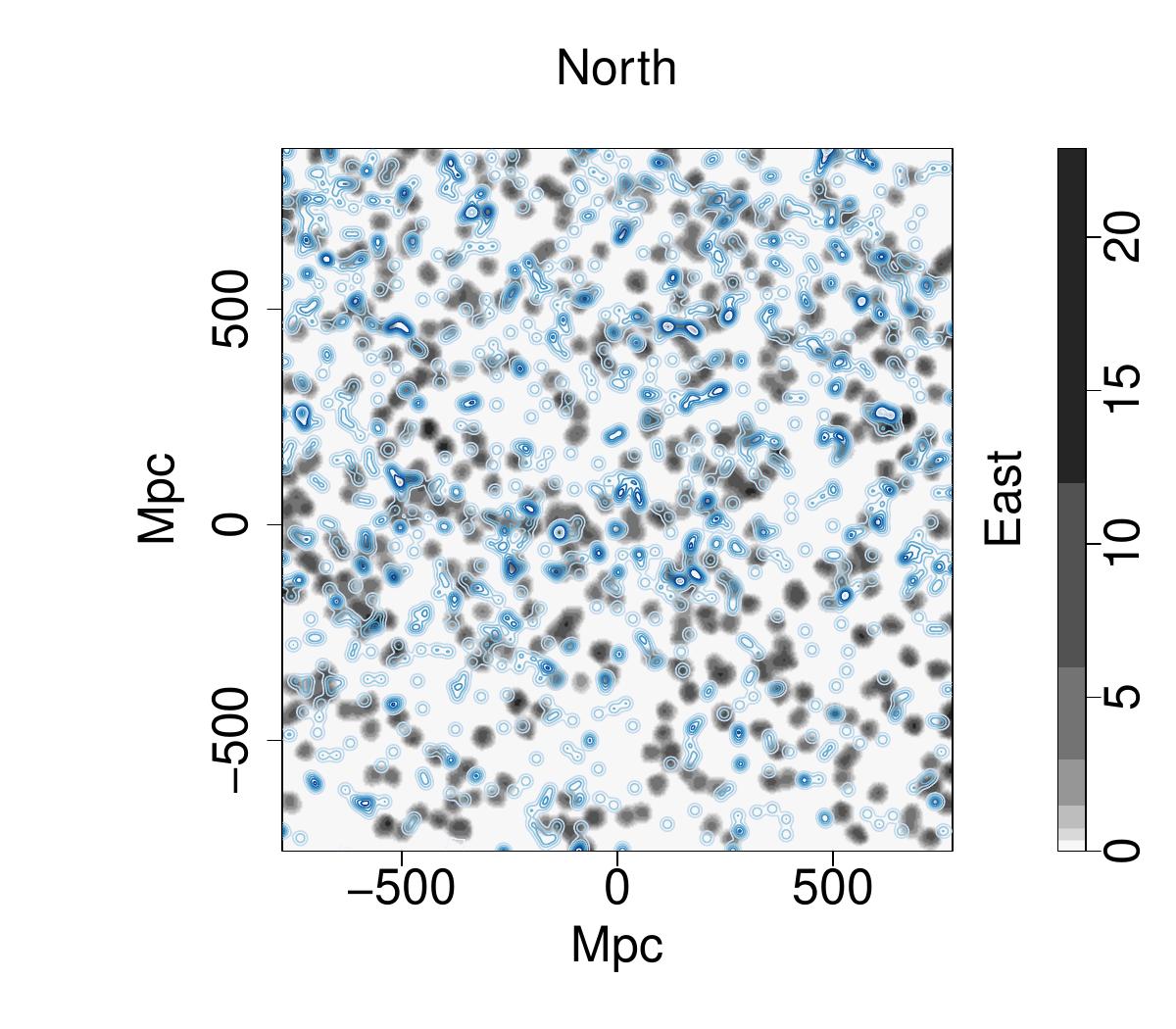}
\caption{}
\label{fig:GA_and_quasars}
\end{subfigure}
\hfill
\begin{subfigure}[b]{0.475\textwidth}
\centering
\includegraphics[width=\textwidth]{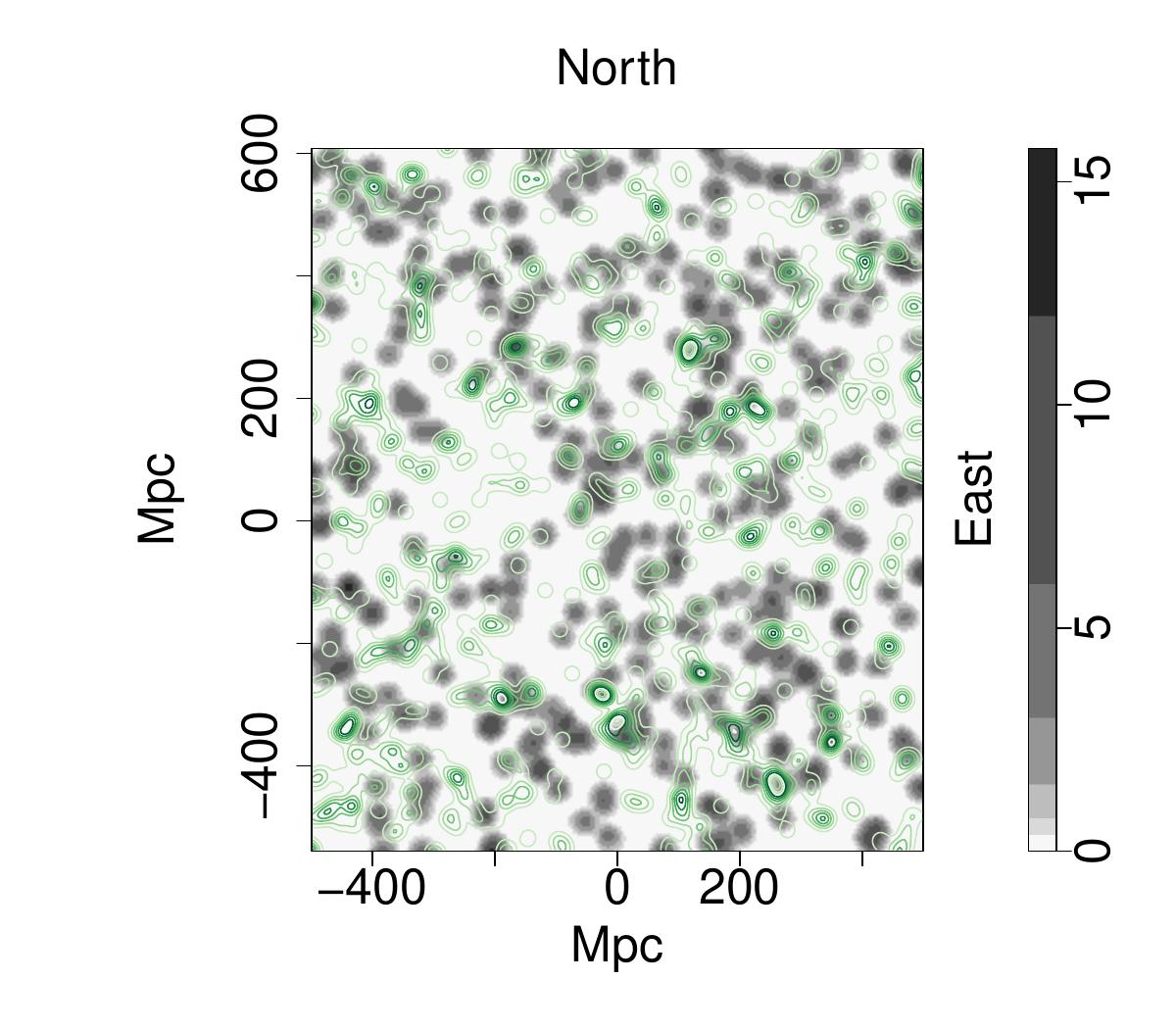}
\caption{}
\label{fig:BR_and_DESI}
\end{subfigure}
\caption{\small (a) Density distribution of the flat-fielded Mg~{\sc II}
  absorbers in the GA field in the redshift slice $z=0.802 \pm 0.060$
  represented by grey contours which have been smoothed using a Gaussian
  kernel of $\sigma = 11$~Mpc and flat-fielded with respect to the background
  probes (background quasars). The blue contours represent the \emph{field}
  quasars from DR16Q --- i.e., the quasars that are in the same field and
  redshift slice as the Mg~{\sc II} absorbers --- which have been restricted
  to $i \leq 20.0$ and have also been smoothed with a Gaussian kernel of
  $\sigma = 11$~Mpc. The GA can be seen stretching $\sim 1$~Gpc across the
  field at the tangent-plane $y$-coordinate $\sim 0$.  This figure
  corresponds to Figure~14(a) in the GA paper. (b) Density distribution of
  the flat-fielded Mg~{\sc II} absorbers in the BR field in the redshift
  slice $z=0.802 \pm 0.060$ represented by grey contours which have been
  smoothed using a Gaussian kernel of $\sigma = 11$~Mpc and flat-fielded with
  respect to the background probes (background quasars).  The green contours
  represent the DESI clusters that are in the same field and redshift slice
  as the Mg~{\sc II} absorbers, which have been restricted to a richness
  limit $R \geq 22.5$, and have also been smoothed with a Gaussian kernel of
  $\sigma = 11$~Mpc. The BR can be seen centred approximately on
  tangent-plane $x, y$ coordinates $\sim(50, 300)$. In both figures, the key
  on the right-hand side shows the relative density of the Mg~{\sc II}
  absorbers, which are increasing by a factor of two.}
\label{figs:independent_corroboration}
\end{figure*}

We used coloured contours superimposed on the Mg~{\sc II} images to visually
inspect the relationship between the independent datasets.  In
Figures~\ref{fig:GA_and_quasars} and \ref{fig:BR_and_DESI} we are seeing two
examples of independent corroboration of the Mg~{\sc II} with the DR16Q field
quasars and the DESI clusters, respectively. Visually, the coloured contours
show a tendency to follow the grey contours, and since the quasars and DESI
clusters are independent sources, they both provide independent corroboration of the GA and BR.

\subsection{The Proximity of the BR to the GA}

The BR discovery is particularly interesting as it was made directly after
the discovery of the GA but using the newer Anand21 Mg~{\sc II} catalogues
instead of the older Z\&M catalogues. The BR was first apparent when looking
at the GA field with the Anand21 database; just to the north of the GA was an
interesting ring shape, which we then investigated further.

On the sky, the GA and the BR are separated by only about $12^\circ$, and
they both lie in exactly the same redshift slice $z=0.802 \pm 0.060$. 
Figure~\ref{fig:BR_and_GA_night_sky} is an approximate projection of the BR and GA
absorber members imprinted onto a night sky image taken from Stellarium.

The direct discovery of a second uLSS, the BR, in the same cosmological
neighbourhood as the first discovery, the GA, leads to the question of
whether we would expect to find many more interesting, huge structures with
circular or other conic-section morphologies.  Perhaps, a more important
question would be to determine \emph {a priori} (i) whether $\Lambda$CDM can
explain the occurrence of such huge structures, and (ii) whether, with the CP
assumed, we should expect to find no other, or many more, of these types of
structures.

\subsection{Hints of an Extended GA}

\begin{figure}[!h]
\centering\includegraphics[width=2.5in]{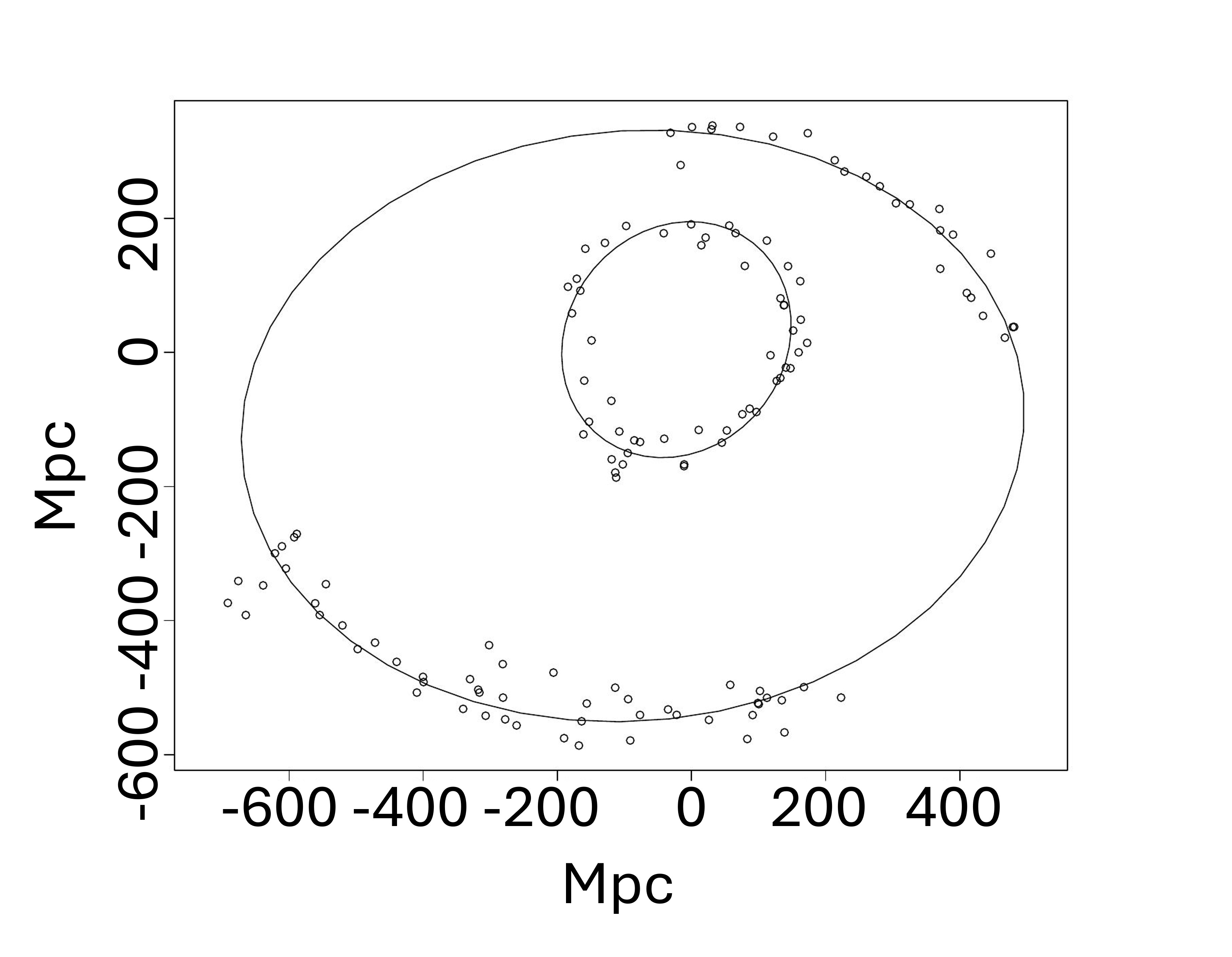}
\caption{The GA points from Z\&M (along the bottom) plotted with the BR
  points from Anand21 (in the middle). The thin filament to the north was
  also detected in the newer Anand21 catalogues. With the fitted ellipses it
  appears that the interesting thin filament could be a continuation of the
  GA, and if so, then this Giant Ring would encompass the BR. }
\label{fig:GA_and_BR_ellipse}
\end{figure}

In the BR field of the Anand21 databases there was also an interesting thin
filament which appears to be a potential continuation of the GA. In
Figure~\ref{fig:GA_and_BR_ellipse} we have added the GA points from the Z\&M
catalogues on the same plot as the BR points from the Anand21 catalogues. In
addition we have added the points of the interesting thin filament that
appeared in the Anand21 catalogues. With fitted ellipses it can be seen that
if (with more data) this interesting filament was a continuation of the GA, then the GA, or
Giant Ring (GR) would then encompass the BR.

\subsection{The 3D distribution of the BR}

For the BR, we define a 3D coordinate system so that the distribution of
Mg~{\sc II} absorber points in BR can be projected onto different planes,
allowing for the investigation of the BR at different viewing angles.

\begin{figure*}
\centering
\begin{subfigure}[b]{0.475\textwidth}
\centering
\includegraphics[width=\textwidth]{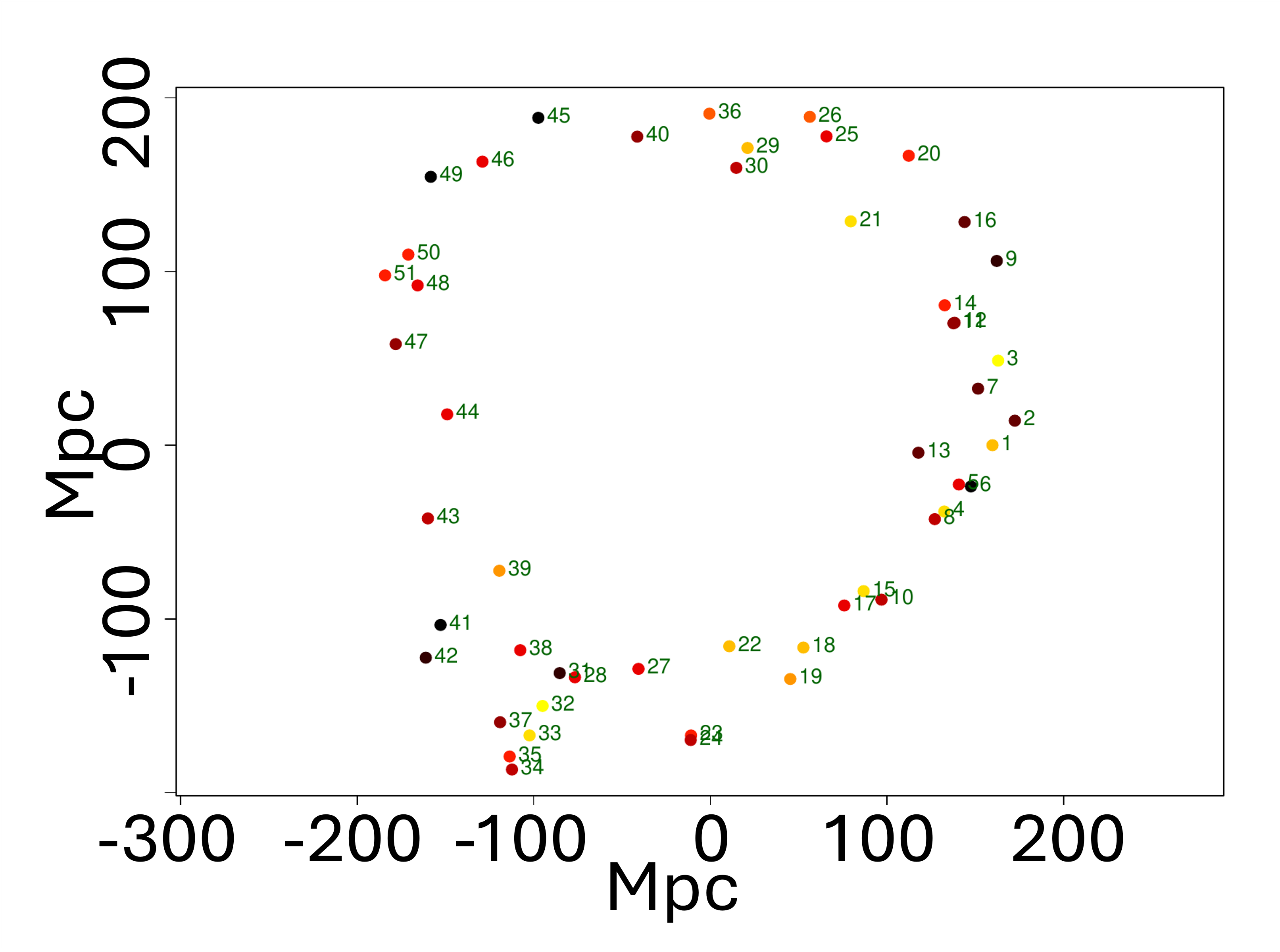}
\caption{ }
\label{subfig:BR_orig_proj}
\end{subfigure}
\hfill
\begin{subfigure}[b]{0.475\textwidth}
\centering
\includegraphics[width=\textwidth]{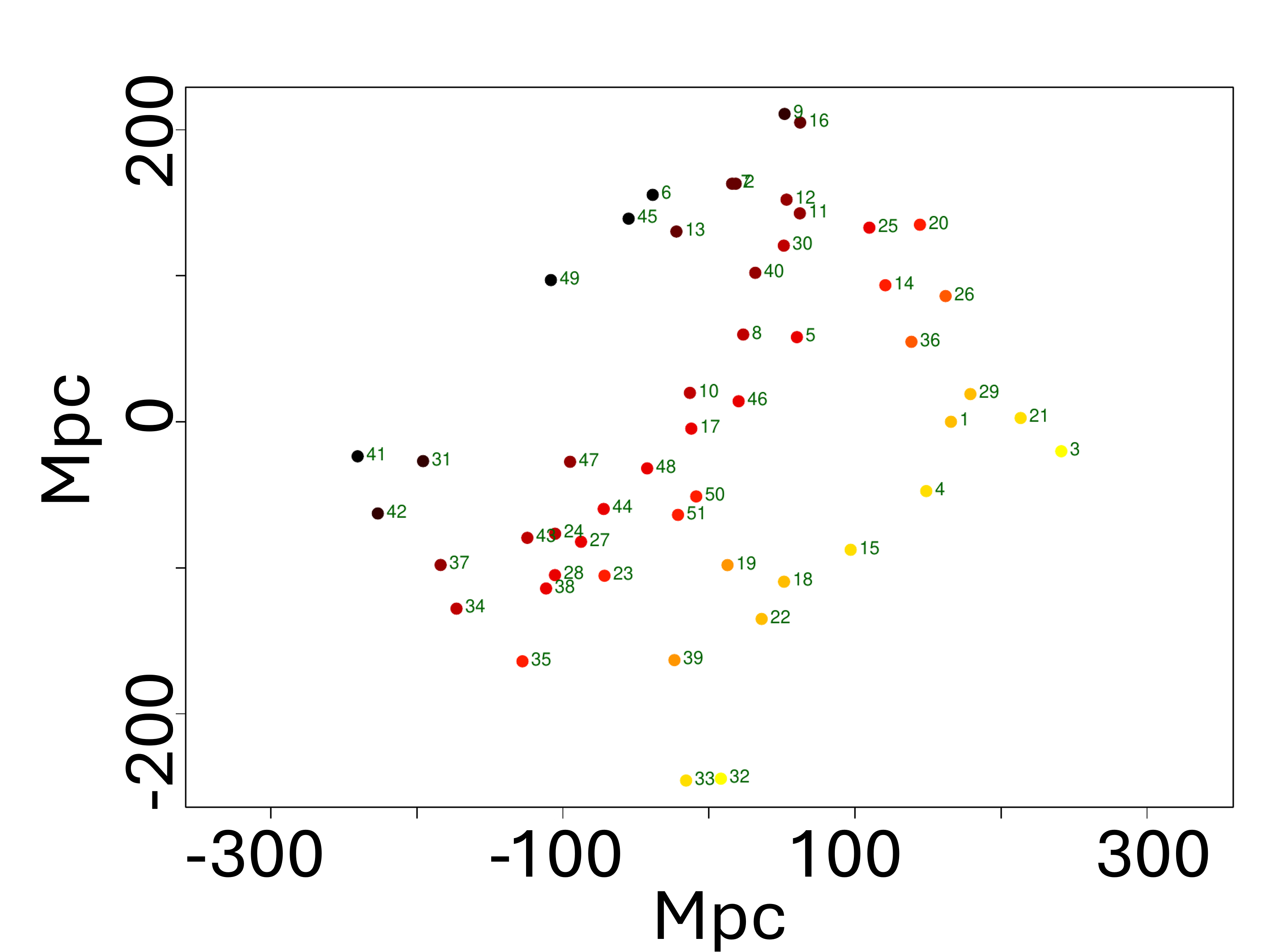}
\caption{}
\label{subfig:BR_proj_2}
\end{subfigure}
\caption{\small (a) The BR absorber members projected onto a plane that
  resembles the original line-of-sight projection.  (b) The BR absorber
  members projected onto a plane that resembles viewing the BR from a
  south-east direction. In this figure, the BR appears to resemble a coil
  shape. In both figures, the colours represent the redshifts of the
  absorbers, with nearer-$z$ absorbers being lighter in colour and further-$z$
  absorbers being darker. The small numbers
  paired with each data point indicate their unique ID number. }
\label{figs:BR_proj}
\end{figure*}

In Figure~\ref{subfig:BR_orig_proj} we are seeing the BR projected onto a
plane, where the 3D coordinate system has been defined to resemble the
original line-of-sight projection. The absorber points are numbered $1-51$,
and these values are fixed to their absorber points through all rotations of
the plane. The colours represent the redshift distribution, with nearer-$z$
absorbers being lighter in colour and further-$z$ absorbers being darker.

In Figure~\ref{subfig:BR_proj_2} we are next seeing the BR absorber
points projected onto a plane that corresponds to viewing the BR from the
south-east direction. From this viewing angle, the BR appears to resemble a
coil shape. There also appear to be three distinct redshift bands, which were
similarly seen from a side-on viewing angle (see the BR paper).

A 3D visualisation tool was also used to move around the BR absorber points
to investigate the 3D structure. Using this tool also supported the
coil-nature of the BR and it gave an impression of the near-$z$ absorbers
`looping' into the central-$z$ absorbers. The central-$z$ absorbers then
contained the majority of the BR absorbers and created the main ring shape in
a thin, flat region. The far-$z$ absorbers showed less continuity, with the
appearance resembling a broken ring shape.

\section{Discussion and Conclusions}

We have reviewed two recently-discovered uLSSs: the Giant Arc and the Big
Ring (GA and BR). Both were discovered visually as prominent overdensities in
Mg~{\sc II} images, and subsequently supported by a range of statistical
assessments and by independent corroboration from DR16Q quasars and DESI
clusters. Both exceed the Yadav estimated upper-limit to the scale of
homogeneity and add to an accumulating list of uLSSs that challenge the
Cosmological Principle (CP) in this respect. We note that interpretations of
the CP vary, however.

The GA and BR have intriguing morphologies, as indicated by their
names. There is even a hint that the GA could extend into a Giant Ring that
envelops the BR. While the BR appears entirely as a ring in projection on the
sky, in 3D visualisations it appears to coil into and out of a central flat
ring (which contains most of the member absorbers).

Extreme-value analysis and cosmological simulations have shown that some
individual occurrences of uLSSs may be consistent with the CP and with
$\Lambda$CDM, but it is not obvious that the entire accumulating set of uLSSs
will be consistent.

Given their morphologies, perhaps the GA and BR in particular require an
explanation outside $\Lambda$CDM. One possibility might be cosmic strings,
which have become of topical interest in recent work \cite{Ahmed2023,
  Cyr2023, Ellis2023, Gouttenoire2023, Jiao2023, Peebles2023, Sanyal2022,
  Wang2023}. For the BR specifically, we noted its similarity in projection
and radius to an individual baryonic acoustic oscillation (BAO)
\cite{Tully2023, Einasto2016, Planck2015, Anderson2014, Eisenstein2005}.
However, given the non-spherical, coiling nature of the BR, and its actually
being larger than a BAO ($r \sim 200$~Mpc compared with $r \sim 150$~Mpc for
a BAO), the BR having an origin in BAOs is probably ruled out.

Unexpected and apparently anomalous discoveries in cosmology, such as uLSSs
in general and the GA and BR in particular, may be indicating a route to
further understanding and to refinements of the standard model. It may be
productive to see where they lead.

\vskip6pt

\ack{Our data has depended on the publicly-available Sloan Digital Sky Survey
  quasar catalogue and the corresponding Mg~{\sc II}
  catalogues. \footnote{For the Anand21 Mg~{\sc II} catalogues, see
  https://wwwmpa.mpa-garching.mpg.de/SDSS/MgII/. For the Z\&M Mg~{\sc II}
  catalogues, see
  https://www.guangtunbenzhu.com/jhu-sdss-metal-absorber-catalog.}  AML was
  supported by a UCLan/JHI PhD studentship.}
    

\end{document}